\documentclass[12pt,epsf]{article}
\usepackage{graphicx}
\usepackage{epsfig}
\usepackage{amsmath}
\usepackage{subfigure}
\usepackage{slashbox}
\usepackage{subfigure}
\begin{document}

\def\a{\alpha}
\def\b{\beta}
\def\c{\varepsilon}
\def\d{\delta}
\def\e{\epsilon}
\def\f{\phi}
\def\g{\gamma}
\def\h{\theta}
\def\k{\kappa}
\def\l{\lambda}
\def\m{\mu}
\def\n{\nu}
\def\p{\psi}
\def\q{\partial}
\def\r{\rho}
\def\s{\sigma}
\def\t{\tau}
\def\u{\upsilon}
\def\v{\varphi}
\def\w{\omega}
\def\x{\xi}
\def\y{\eta}
\def\z{\zeta}
\def\D{\Delta}
\def\G{\Gamma}
\def\H{\Theta}
\def\L{\Lambda}
\def\F{\Phi}
\def\P{\Psi}
\def\S{\Sigma}

\def\o{\over}
\def\beq{\begin{eqnarray}}
\def\eeq{\end{eqnarray}}
\newcommand{\lsim}{\raisebox{0.6mm}{$\, <$} \hspace{-3.0mm}\raisebox{-1.5mm}{\em $\sim \,$}}
\newcommand{\gsim}{\raisebox{0.6mm}{$\, >$} \hspace{-3.0mm}\raisebox{-1.5mm}{\em $\sim \,$}}

\newcommand{\vev}[1]{ \left\langle {#1} \right\rangle }
\newcommand{\bra}[1]{ \langle {#1} | }
\newcommand{\ket}[1]{ | {#1} \rangle }
\newcommand{\EV}{ {\rm eV} }
\newcommand{\MeV}{ {\rm MeV} }
\newcommand{\GeV}{ {\rm GeV} }
\newcommand{\TeV}{ {\rm TeV} }
\def\diag{\mathop{\rm diag}\nolimits}
\def\Spin{\mathop{\rm Spin}}
\def\SO{\mathop{\rm SO}}
\def\O{\mathop{\rm O}}
\def\SU{\mathop{\rm SU}}
\def\U{\mathop{\rm U}}
\def\Sp{\mathop{\rm Sp}}
\def\SL{\mathop{\rm SL}}
\def\tr{\mathop{\rm tr}}

\def\IJMP{Int.~J.~Mod.~Phys. }
\def\MPL{Mod.~Phys.~Lett. }
\def\NP{Nucl.~Phys. }
\def\PL{Phys.~Lett. }
\def\PR{Phys.~Rev. }
\def\PRL{Phys.~Rev.~Lett. }
\def\PTP{Prog.~Theor.~Phys. }
\def\ZP{Z.~Phys. }

\newcommand{\KEV}{ {\rm keV} }
\newcommand{\MEV}{ {\rm MeV} }
\newcommand{\GEV}{ {\rm GeV} }
\newcommand{\TEV}{ {\rm TeV} }

\def\Z{\mathcal{Z}}
\def\W{\Omega}

\baselineskip 0.7cm

\begin{titlepage}

\begin{flushright}
IPMU10-0192\\
UT-10-20
\end{flushright}

\vskip 1.35cm
\begin{center}
{\large \bf
Discovery Potential for
Low-Scale Gauge Mediation at Early LHC
}
\vskip 1.2cm
Eita Nakamura and Satoshi Shirai
\vskip 0.4cm

{\it
Department of Physics, University of Tokyo, 
Tokyo 113-0033,
Japan\\
IPMU,
University of Tokyo, 
Chiba 277-8586, 
Japan\\
}

\vskip 1.5cm

\abstract{
Low-scale gauge-mediated supersymmetry(SUSY)-breaking (GMSB) models with gravitino mass $m_{3/2}<16$ eV are attractive, 
since there are no flavor and cosmological problems.
In this paper, we thoroughly study the collider signal in the case that the next-to-lightest SUSY particle is the bino or slepton
and investigate the discovery potential of the LHC.
Our result is applicable to a wider class of GMSB models other than the minimal GMSB models
and we pay particular attention to realistic experimental setups.
We also apply our analysis to the minimal GMSB models with a metastable SUSY-breaking vacuum and we show, by requiring sufficient stability
of the SUSY-breaking vacuum, these models can be tested at an early stage of the LHC.
}
\end{center}
\end{titlepage}

\setcounter{page}{2}

\section{Introduction}
Supersymmetric (SUSY) standard model (SSM) is the most promising candidate for the physics beyond the standard model (SM).
Among proposed models of SUSY-breaking mediation mechanisms, the gauge-mediated SUSY-breaking 
(GMSB) model \cite{Giudice:1998bp}  is very attractive, 
since there are no SUSY flavor problems.
In addition, if the gravitino ($\tilde{G}$) is lighter than 16 eV, there are no cosmological gravitino problems \cite{Viel:2005qj}.
Such a light gravitino plays an important role in collider physics, e.g. at the LHC.
Once SUSY particles are produced at the LHC, they successively decay into lighter particles and finally
into the next-to-lightest SUSY particle (NLSP).
The NLSP then decays into a gravitino, which is the lightest SUSY particle (LSP).
In a typical GMSB model, the NLSP is the bino-like neutralino or righted-handed slepton, since their gauge interactions are weak.
Their dominant decay modes are $\tilde{\chi}^0_1 \to \gamma + \tilde{G}$ and $\tilde{\ell} \to \ell + \tilde{G}$, respectively.
Thus the LHC signal is multi-photon+missing energy, or multi-lepton+missing energy.
For both signals, there are tiny SM backgrounds.
Thus, the low-scale GMSB scenario is easier to be tested at the LHC compared to other SUSY models.

There are many studies for the discovery potential of the LHC for low-scale GMSB models at the LHC \cite{Baer:1998ve,Baer:2000pe,CSCnote,Harper:2009iw,Ludwig:2010kt}.
Although these studies are basically based on the minimal GMSB models,
general GMSB models have wider parameter space.
In this paper, we investigate the LHC discovery potential for more generic parameter space than the one of the minimal GMSB models.
We perform a comprehensive analysis using all known discovery modes
and pay particular attention to realistic experimental setups.

From a theoretical viewpoint, such a light-gravitino scenario tends to suffer from a serious problem: 
instability of the SUSY-breaking vacuum.
There is a strong upper bound on the SUSY-particle masses for the minimal GMSB models
in which the SUSY-breaking vacuum is metastable \cite{Hisano:2007gb}.
We apply our analysis of the discovery potential to such GMSB models
and find that these models can be tested at an early stage of the LHC.
Especially, if the number of messengers is one, it is possible to test at a very early stage: 1 ${\rm fb}^{-1}$ at $\sqrt{s}=7$ TeV.

\section{Model Setup}

In this section, we illustrate the setup of the models we consider in this paper.

We first consider a simple GMSB model, where a SUSY-breaking field $S$
couples to $N_5$ pairs of messenger chiral superfields, $\P$ and
$\bar{\P}$, which transform as ${\bf 5}$ and $\bar{\bf 5}$ under the gauge group
$\SU(5)_{\rm GUT}$.
The simplest form of the coupling of the messengers and the SUSY-breaking field is
\beq
W \;=\; \sum_{i=1}^{N_5}(y_i S+M_i) \Psi_i{\bar \Psi}_i,
\eeq
where $M_i$ is the messenger mass and $y_i$ is a Yukawa coupling constant and assumed to be less than ${\cal O}(1)$.
By assumption, the SUSY-breaking chiral field $S$ develops an $F$-term vacuum expectation value (VEV) $\langle S \rangle = \theta^2 F$,
 which is related to the gravitino mass as $|F| = \sqrt{3} m_{3/2} M_{PL}$, assuming that
the SUSY breaking is dominated by $F$.
The condition that $M_i^2>y_iF$ should be satisfied, otherwise the messenger scalars are tachyonic.

In the minimal GMSB models, the MSSM gaugino masses are generated from loop
diagrams of the messengers. At the one-loop level, the gaugino masses are
given by
\begin{equation}
M_{a} \;=\; \frac{\alpha_a}{4\pi}\Lambda_g(1+{\cal O}(x_i^2)),
\label{eq:gaugino_mass2}
\end{equation}
where we have defined
\beq
\Lambda_g = \sum_{i=1}^{N_5}\frac{y_i F}{M_i},
\eeq
and $x_i=y_i F/M_i^2$.
Here, $a (= 1,2,3)$ labels the SM gauge group U(1), SU(2) and SU(3), 
and we use the normalization $\alpha_1=5 \alpha _{\rm EM}/(3 \cos^2\theta_{W})$.
The scalar masses arise at the  two-loop level, and are given by
\begin{equation}
m^2_{\phi_i}\;=\;\Lambda_s^2 \sum_a \left(\frac{\alpha_a}{4\pi}\right)^2 C_a (i)(1+{\cal O}(x_i^2)),
\label{eq:scalar_mass2}
\end{equation}
where
\beq
\Lambda_s^2 = 2\sum_{i=1}^{N_5}\left|\frac{y_i F}{M_i}\right|^2
\eeq
and $C_a(i)$ are the Casimir invariants for the visible particles $\phi_i$ 
($C_1(i) = 3Y_i^2/5$).
\footnote{We assume that the gaugino and scalar masses are parametrized by one parameter, $\Lambda_g$ and $\Lambda_s$, respectively.
In general case, this universality is not maintained since the GUT is broken at the messenger scale and there are two types of the messengers,
lepton and quark-type messengers.
However, once we assume the both masses and yukawa couplings of the lepton and quark-type messengers are identical at the GUT scale,
such  universality is  maintained.
}
Each $x_i$ is bounded as $x_i<1$ for the messengers not to become tachyonic, and then the corrections of ${\cal O}(x_i^2)$ 
are typically small and we ignore these corrections in the following analysis.
We see that $m_{\f_i} \simeq M_a = {\cal O}(1)~\TEV$ is realized for $\Lambda_s\simeq\Lambda_g={\cal O}(100)~\TEV$.
To realize $m_{3/2}\lsim 16$ eV, the mass of the messenger must be $M_i={\cal O}(100)$ TeV.

While conventional studies on LHC physics of GMSB models have been performed with the above setup
of minimal GMSB models, general GMSB models exhibit different patterns
of mass generation. To be more precise,
in the minimal GMSB models, the two scales $\Lambda_g$ and $\Lambda_s$ have a certain relation.
However, in many examples of GMSB models the relations for $\Lambda_g$ and $\Lambda_s$
are modified.
For example, if the mass matrix of the messenger is complicated, the relation between 
$\Lambda_g$ and $\Lambda_s$ is also complicated \cite{Cheung:2007es}.
Another example is the strongly interacting messenger.
If the messenger has large anomalous dimension, the relation between $\Lambda_g$ and $\Lambda_s$ is deformed.
Therefore, we treat the $\Lambda_g$ and $\Lambda_s$ as independent free parameters in
order that the analysis can be applied to a wider class of low-scale GMSB models.

In the most general setup of a GMSB model \cite{Meade:2008wd}, the pattern of mass generation is further more general than
the one in our setup. However, essential part of our analysis can be also applied
to this general case. We will come back to this discussion in Sec. \ref{general}.

\subsection*{Masses of MSSM Particles}
Since the above
expressions Eqs.~(\ref{eq:gaugino_mass2}) and (\ref{eq:scalar_mass2}) for the SUSY-particle masses are 
given at the messenger scale, one should solve the MSSM renormalization group equation to
obtain the on-shell masses and mixing matrices. 
To calculate the on-shell masses and other physical parameters, 
we use the program ISAJET 7.72 \cite{ISAJET} slightly modified by the authors.
We adopt 200 TeV as the messenger mass.

The mass spectrum and therefore the low-energy phenomenology depend on the parameter ${\rm tan}\,\beta$,
which is the ratio between the up-type and down-type Higgs VEVs.
In most part of our analysis, we use the values $\tan\beta= 10$ and 40 to represent the low-${\rm tan}\,\beta$
and high-${\rm tan}\,\beta$ cases. The effect of varying ${\rm tan}\,\beta$ is discussed in Sec. \ref{minimal}.

\begin{figure}
\subfigure[$\tan\beta=10$]{
\includegraphics[clip, width=0.5\columnwidth]{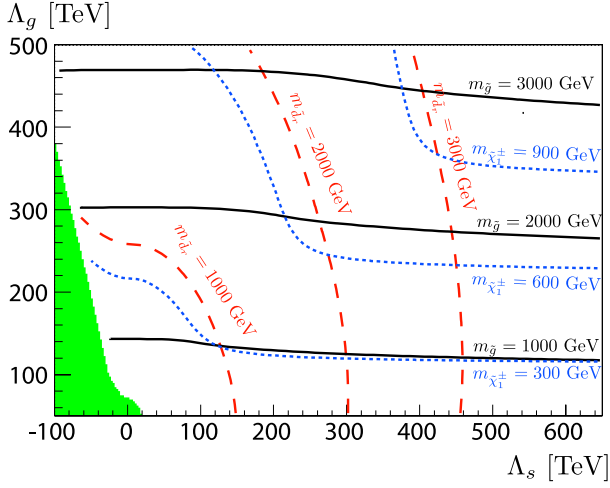}
}
\subfigure[$\tan\beta=10$]{
\includegraphics[clip, width=0.5\columnwidth]{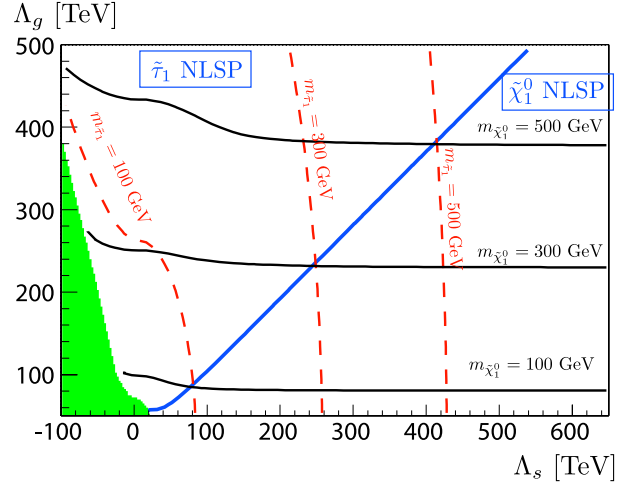}
}\\
\subfigure[$\tan\beta=40$]{
\includegraphics[clip, width=0.5\columnwidth]{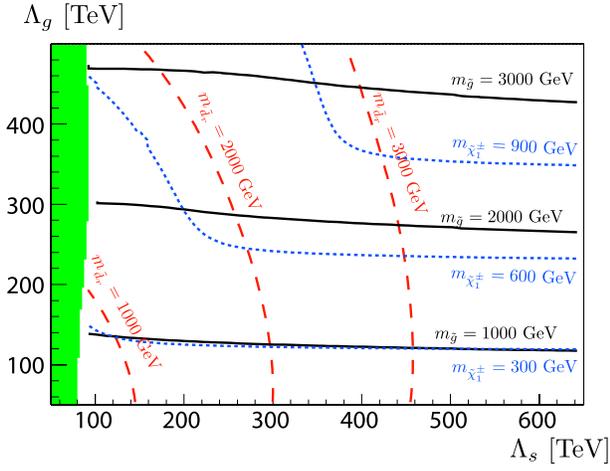}
}
\subfigure[$\tan\beta=40$]{
\includegraphics[clip, width=0.5\columnwidth]{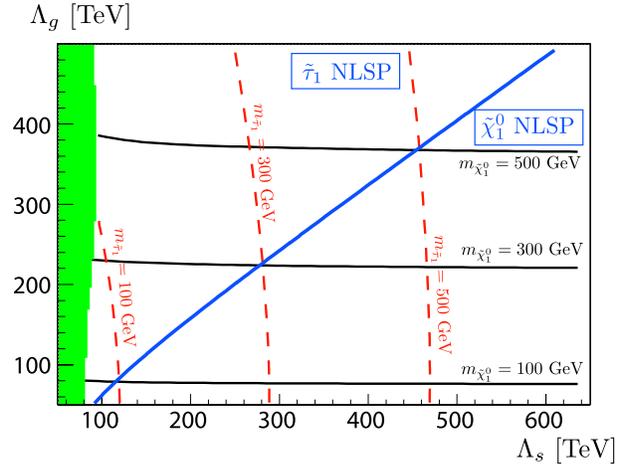}
}
\caption{Contour plot of the masses of the MSSM particles. In the left figures (a) and (c), the squark, gluino and wino masses are
shown. In the right figures (b) and (d), the lighter stau mass and the lightest neutralino mass are shown.
Figures (a) and (b) shows masses for the case with
${\rm tan}\,\beta=10$ and (c) and (d) for ${\rm tan}\,\beta=40$.
The green region in each figure shows that a correct electroweak symmetry breaking cannot be achieved there.
In figures (b) and (d), the blue line shows the points where the lighter stau and the lightest neutralino
have the same mass.}
\label{fig:mass}
\end{figure}

In Fig. \ref{fig:mass}, we illustrate the masses of SUSY particles. For the case with ${\rm tan}\,\beta=10$,
we extend the range of $\Lambda_s$ to negative values, where we define ``$\Lambda_s^2$'' to be ${\rm sign}(\Lambda_s)|\Lambda_s|^2$.
This is because generically we may well
have negative scalar masses at the messenger mass scale, e.g. with the $D$-term contribution.
Since the parameter region is not excluded either theoretically or experimentally,
we include the region for completeness. 
For $\Lambda_g$, the lower value is experimentally excluded as we illustrate in later figures.
The green region in each figure is the region
in which the correct electroweak symmetry breaking cannot be achieved.
In the left figures (a) and (c), the masses of the squark, gluino, and lightest chargino
are illustrated. As we see below, these particles are important for the SUSY production at the LHC and
their masses often determine the cross sections and thus discovery reaches. The masses of stau and the
lightest neutralino are drawn in the right figures (b) and (d). These are the lightest SUSY particles in
the models and we also draw a boundary curve where the NLSP changes. In the right side of this boundary,
the neutralino is the NLSP and in the left side the stau is the NLSP. Thus, this boundary corresponds to
a change in searching channel, from diphoton + missing to multi-lepton + missing.

\section{LHC Signature and Discovery Potential}

\subsection{LHC Signature}
In low-scale GMSB models, the gravitino plays an important role at the LHC.
The produced SUSY particles successively decay into the NLSP and
finally, the NLSP decays into the gravitino.
The decay length of the NLSP is roughly given as
\begin{equation}
c\tau\sim 20~{\mu}{\rm m} \left(\frac{m_{3/2}}{1~{\rm eV}} \right)^2\left(\frac{m_{\rm NLSP}}{100~{\rm GeV}}  \right)^{-5}.
\end{equation}
Therefore, if the gravitino is light the decay of the NLSP occurs promptly.
In the present setup, the NLSP is the bino-like neutralino or the righted-handed slepton.
Their main decay modes are $\tilde{\chi}^0_1 \to \gamma + \tilde{G}$ and $\tilde{\ell} \to \ell + \tilde{G}$, respectively.
Thus the LHC signal is multi-photon+missing energy or multi-lepton+missing energy.

\subsection{Analysis}
In the following analysis, we set the gravitino mass $m_{3/2} = 16$ eV.
We have used the programs 
ISAJET 7.72 \cite{ISAJET} to generate the MSSM mass spectrum and decay table
and Herwig 6.510 \cite{HERWIG6510} to generate SUSY events at the LHC.
For detector simulation, we have used AcerDet 1.0\cite{RichterWas:2002ch} slightly modified by the authors.

\subsubsection{Detection Efficiency and Misidentification}
In the fast simulation, the detection efficiencies of a photon or a lepton which passes
a certain isolation criteria are $100\%$.
However in multi-lepton or photon signals, the reduction of the number of signal events originated  from
misidentifications of these particles is important.
In our simulation, we include the fake rates of the leptons, photons and jets.
The fake rates we have used in the analysis are shown in Table \ref{tb:fake}.

\begin{table}[htbp]
\caption{Fake rates.}
\label{tb:fake}
\begin{tabular}{|c|c|c|c|c|c|c|c|c|c|}
\hline
&$j\to$ e & $j\to \mu$&  $j \to \gamma$ &$\tau$ $\to j$  & $j\to\tau$& $e \to j$ &$\mu \to j$  & $e \to \gamma$   & $\gamma \to j$ \\ \hline
SUSY&0.3 \%  &0.3 \%  &0.02 \% & 60 \%&1 \% & 27 \% & 30 \% &3 \% & 20 \% \\ \hline
SM BG&0.3 \% &0.3 \% &0.02 \% &20 \% & 1 \% &0 \% & 0 \% & 3 \%& 0 \% \\
\hline
\end{tabular}
\end{table}

\subsubsection{SM Background}
In the present model, high energy leptons or photons accompany the SUSY events at the LHC.
Thus, there are few backgrounds from the QCD events.
The main background comes from $t{\bar t}$ and gauge boson production events.
To estimate the background, we have used the programs MC@NLO 3.42\cite{Frixione:2002ik}
(for $t\bar{t},WW,WZ$ and $ZZ$),
Alpgen 2.13\cite{Mangano:2002ea} (for $Wj,Zj$ and $W/Z+b\bar{b}/t\bar{t}$),
MadGraph 4.1.44 \cite{Alwall:2007st} (for $t\bar{t}/W/Z+\gamma/\gamma\gamma$), and
Pythia 6.4 \cite{Sjostrand:2006za} (for $\gamma\gamma$).

In contrast to the SUSY signal events, we assume that photon and lepton detection efficiencies are 100 \%
for the SM backgrounds (see Table \ref{tb:fake}).
The misidentified jets from the SM backgrounds can be significant for multi-lepton and multi-photon modes.
Therefore, for the fake rate from jets to leptons and photons, the SM backgrounds are treated in 
the same way as the signal events.
Therefore, our estimates for the event number of SM backgrounds are conservative.

\subsubsection{Event Cuts and Optimization}\label{modes}

To illustrate the discovery potential of the LHC, we calculate the optimized significance of each model point
against the SM backgrounds. To incorporate both the statistical and systematic uncertainties of the backgrounds,
we adopt the method used in Ref. \cite{CSCnote}. The systematic uncertainties of the backgrounds are taken to be
$50\%$ for the QCD multi-jet background and $20\%$ for others.

Given the number of signal events $N_s$, background events $N_b$ with the uncertainty $\delta N_b$,
the significance is given by calculating the convolution of the Poisson distribution with some ``posterior'' distribution
function. As the posterior distribution, we take the gamma distribution as suggested in Ref. \cite{Linnemann}.
The resulting significance $Z_B$ is given by \cite{Linnemann}
\begin{equation}
Z_B=\sqrt{2}{\rm erf}^{-1}(1-2p_B),
\end{equation}
with
\begin{equation}\label{prob}
p_B=\frac{B(N_s+N_b,1+N_b^2/\delta N_b^2,\delta N_b^2/(N_b+\delta N_b^2))}{B(N_s+N_b,1+N_b^2/\delta N_b^2)},
\end{equation}
where ${\rm erf}^{-1}$ is the inverse error function and
\begin{equation}
B(a,b,x)=\int_0^x dt\, t^{a-1}(1-t)^{b-1}
\end{equation}
is the incomplete beta function and $B(a,b)=B(a,b,1)$ is the usual beta function. 
If we take the limit $\delta N_b\to0$, the Eq.~(\ref{prob}) reduces to the probability
in the usual Poisson distribution.
When $N_b\simeq0.1$ and $N_s\lsim10$, $Z_B\simeq N_s$. 
In the case of smaller background $N_b<0.1$, we conservatively take the $N_s$ as the significance.

For each model point, significances, as defined above, are calculated with some sets of kinematical cuts
and a set of cuts which gives the largest significance is selected.
After this optimization, the condition $Z_B>5$ is used as a criterion for discovery.

\subsubsection{Search Modes and Kinematical Cuts}
As we discussed above, the LHC signal of the model is multi-photon $+$ missing or multi-lepton $+$ missing.
In the following, we list the search modes which we investigate and the set of kinematical cuts we used
for the optimization of the significance.

We perform exclusive searches on the number of photons and of leptons. For the photon mode, we analyze
one photon ($1\gamma$) mode and two or more photon ($2\geq\gamma$) mode. For the lepton mode, 
we analyze zero lepton ($0\ell$) mode , one lepton ($1\ell$) mode,
same-sign two leptons (SS$2\ell$) mode, three leptons ($3\ell$) mode and four or more leptons ($4\geq\ell$) mode. 
Each mode is divided into submodes
according to the number of leptons (for the one photon mode) or the number of tau jets (for the lepton modes).
These modes are summarized as:
\begin{equation}
\begin{cases}
\textbf{Photon modes}
\begin{cases}
1\gamma+(1\ell, 2\geq\ell)\\
2\geq\gamma
\end{cases}\\
\text{  }\\
\textbf{Lepton modes}
\begin{cases}
0\ell+(2\tau,\,3\geq\tau)\\
1\ell+(0\tau,\,1\tau,\,2\geq\tau)\\
{\rm SS}2\ell+(0\tau,\,1\tau,\,2\geq\tau)\\
3\ell+(0\tau,\,1\tau,\,2\geq\tau)\\
4\geq\ell
\end{cases}
\end{cases}
\end{equation}

As mentioned in the previous subsection, we use a set of kinematical cuts for optimization of the significance.
More concretely, we prepare steps of cuts for the missing energy, number of jets,
jet $p_{\rm T}$, photon $p_{\rm T}$, lepton $p_{\rm T}$ and tau-jet $p_{\rm T}$ as follows:
\begin{align}
&\{E_{\rm T,miss}>50, >100, >150,>200,>300\}({\rm GeV}),\label{Etmiss}\\
&\{N_{\rm jets}\geq0,\geq1,\geq2,\geq3,\geq4\}\notag\\
&\hspace{30pt}\text{where $p_{\rm T}(j_1)>100$ GeV, $p_{\rm T}(j_{2,3,...})>50$ GeV},\label{Njet}\\
&\{p_{\rm T}(j)\geq 0,>100,>200,>300,>500\}({\rm GeV}),\label{Ptjet}\\
&\{p_{\rm T}(\gamma)>30,>60,>90\}({\rm GeV}),\label{Ptphot}\\
&\{p_{\rm T}(\ell)>10,>20,>30,>50,>70\}({\rm GeV}),\label{Ptlep}\\
&\{p_{\rm T}(\tau)>20,>30,>40,>60,>80\}({\rm GeV}). \label{Pttau}
\end{align}
We show in Table \ref{tab_cut} which set of cuts are used for optimization in each mode.
\begin{table}[h!]
\caption{Kinematical cuts used for each mode.}
\label{tab_cut}
\begin{tabular}{|l|c|c|c|c|c|c|}
\hline
mode & Eq.(\ref{Etmiss}) & Eq.(\ref{Njet}) & Eq.(\ref{Ptjet}) & Eq.(\ref{Ptphot}) & Eq.(\ref{Ptlep}) & Eq.(\ref{Pttau})\\\hline\hline
$1\gamma+1\ell$&$\surd$		&$\surd$&&$\gamma_1$&$\ell_1$			&	\\\hline
$1\gamma+2\geq\ell$&$\surd$	&$\surd$&&$\gamma_1$&$\ell_1,\ell_2$		&	\\\hline
$2\geq\gamma$&$\surd$		&$\surd$&&$\gamma_1,\gamma_2$&			&	\\\hline\hline
$0\ell+2\tau$&$\surd$		&&$j_1,j_2$&&($p_{\rm T}(\ell_1){<}10$ GeV)	&$\tau_1,\tau_2$	\\\hline
$0\ell+3\geq\tau$&$\surd$	&&$j_1,j_2$&&($p_{\rm T}(\ell_1){<}10$ GeV)	&$\tau_1,\tau_3$	\\\hline\hline
$1\ell+0\tau$&$\surd$		&&$j_1,j_2$&&$\ell_1$					&($p_{\rm T}(\tau_1){<}20$ GeV)	\\\hline
$1\ell+1\tau$&$\surd$		&&$j_1,j_2$&&$\ell_1$					&$\tau_1$	\\\hline
$1\ell+2\geq\tau$&$\surd$	&&$j_1,j_2$&&$\ell_1$					&$\tau_1,\tau_2$	\\\hline\hline
SS$2\ell+0\tau$&$\surd$		&&$j_1,j_2$&&$\ell_1,\ell_2$				&($p_{\rm T}(\tau_1){<}20$ GeV)	\\\hline
SS$2\ell+1\tau$&$\surd$		&&$j_1,j_2$&&$\ell_1,\ell_2$				&$\tau_1$	\\\hline
SS$2\ell+2\geq\tau$&$\surd$	&&$j_1,j_2$&&$\ell_1,\ell_2$				&$\tau_1,\tau_2$	\\\hline\hline
$3\ell+0\tau$&$\surd$		&&$j_1,j_2$&&$\ell_1,\ell_3$				&($p_{\rm T}(\tau_1){<}20$ GeV)	\\\hline
$3\ell+1\tau$&$\surd$		&&$j_1,j_2$&&$\ell_1,\ell_3$				&$\tau_1$	\\\hline
$3\ell+2\geq\tau$&$\surd$	&&$j_1,j_2$&&$\ell_1,\ell_3$				&$\tau_1,\tau_2$	\\\hline\hline
$4\geq\ell$&$\surd$		&&$j_1,j_2$&&$\ell_1,\ell_4$				&	\\\hline
\end{tabular}
\end{table}
For example, in the three leptons $+$ two or more tau-jets mode, the cut on the missing energy $E_{{\rm T},{\rm miss}}=|{\bf p}_{\rm T,miss}|$
is selected from the set in Eq. (\ref{Etmiss}), cuts on the first two jets, $j_1$ and $j_2$, are selected from
Eq. (\ref{Ptjet}), cuts on the first and third leptons, $\ell_1$ and $\ell_3$, are selected from
Eq. (\ref{Ptlep}) and cuts on the first and second tau-jets, $\tau_1$ and $\tau_2$, are
selected from Eq. (\ref{Pttau}).

\subsection{LHC Reach}
\begin{figure}
\subfigure[2 photons]{
\includegraphics[clip, width=0.5\columnwidth]{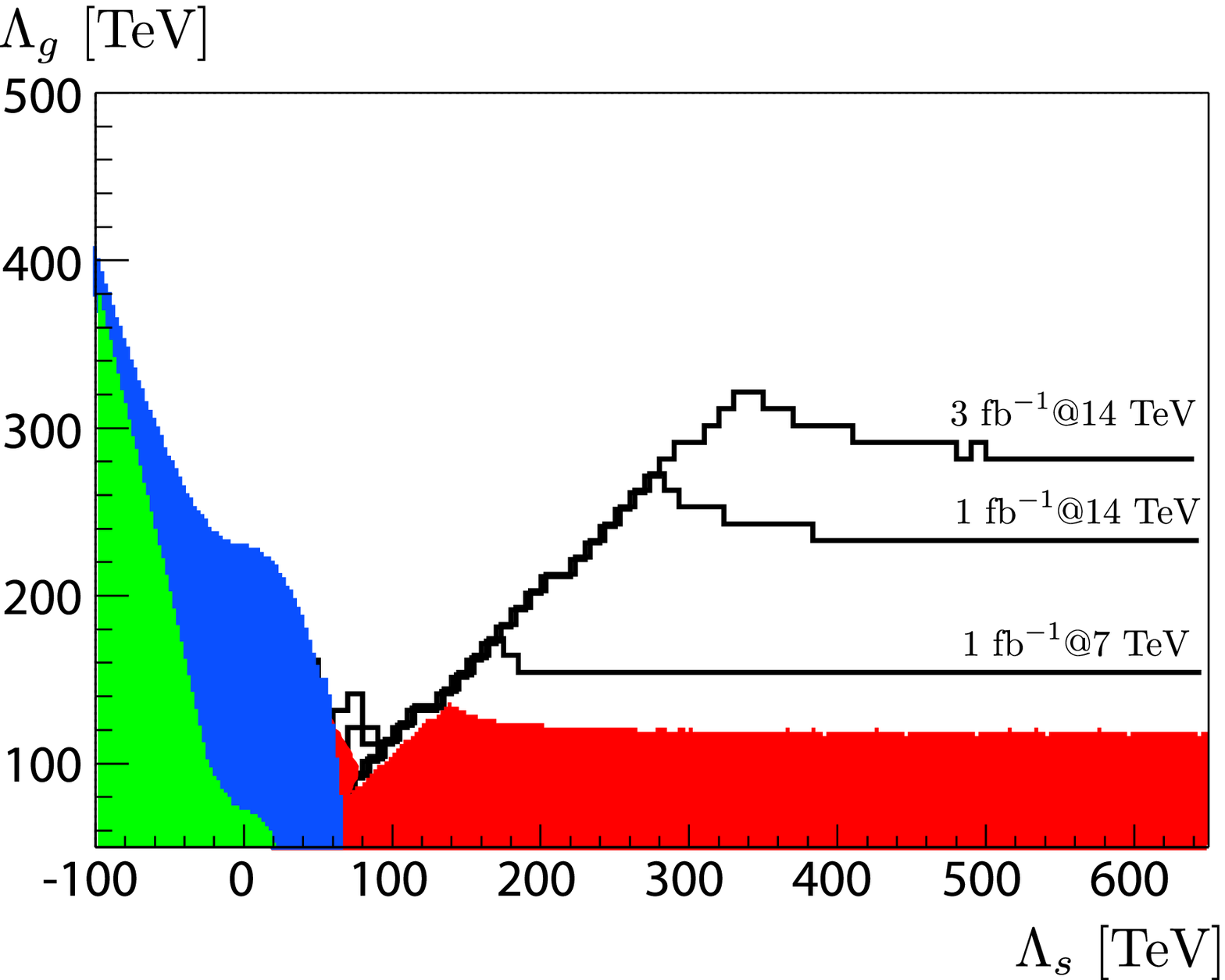}
}
\subfigure[3 leptons]{
\includegraphics[clip, width=0.5\columnwidth]{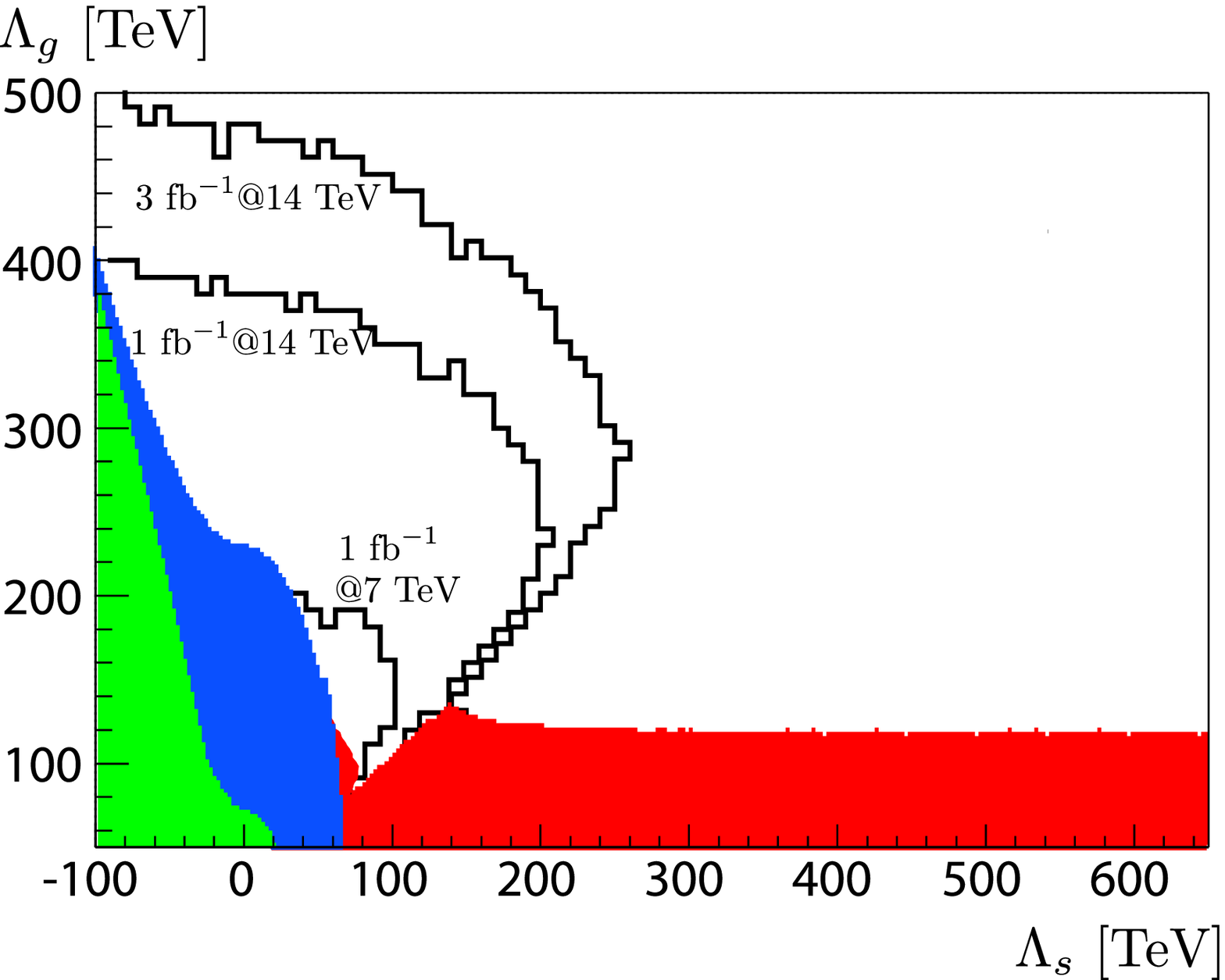}
}\\
\subfigure[4 or more leptons]{
\includegraphics[clip, width=0.5\columnwidth]{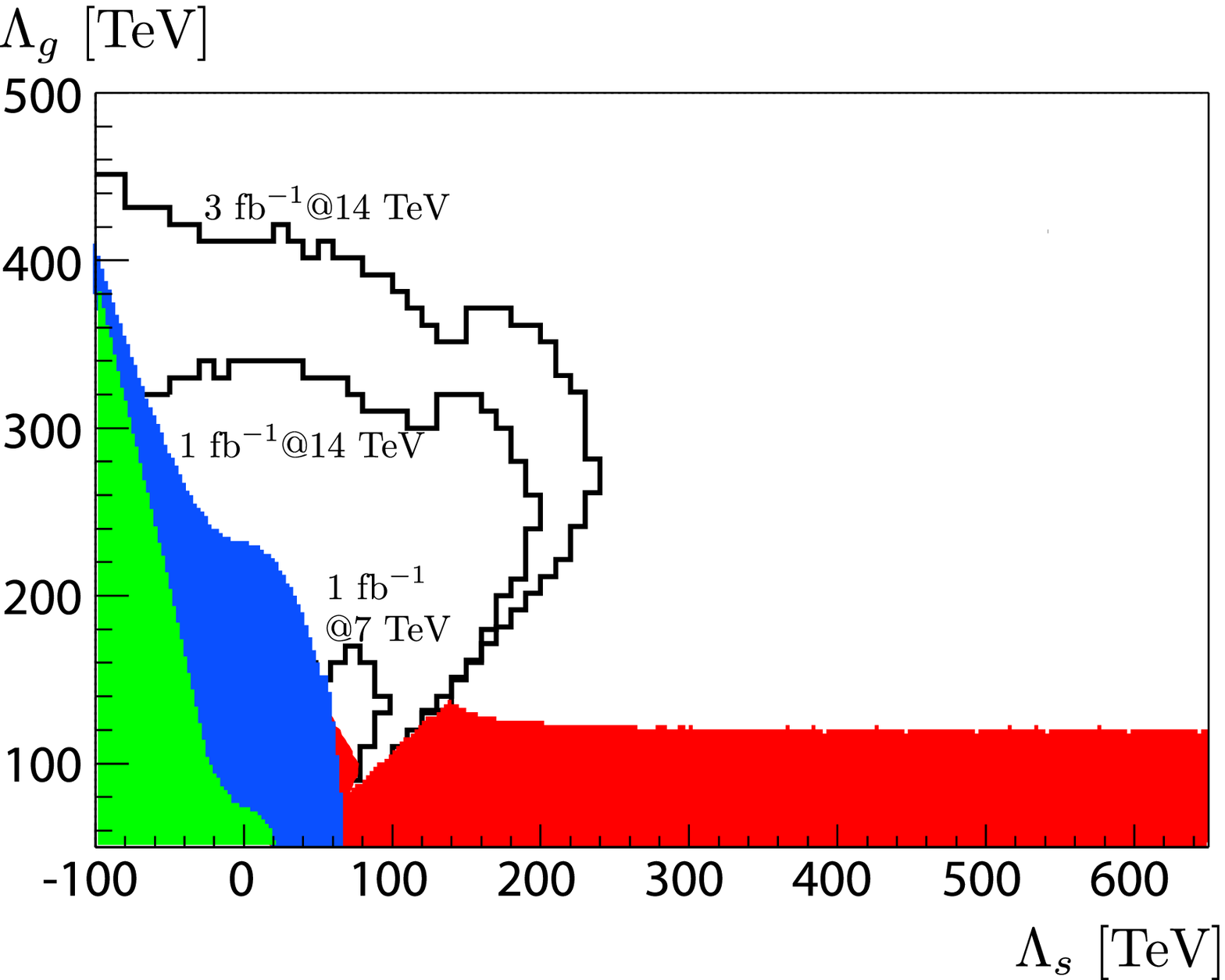}
}
\subfigure[Same-sign 2 leptons]{
\includegraphics[clip, width=0.5\columnwidth]{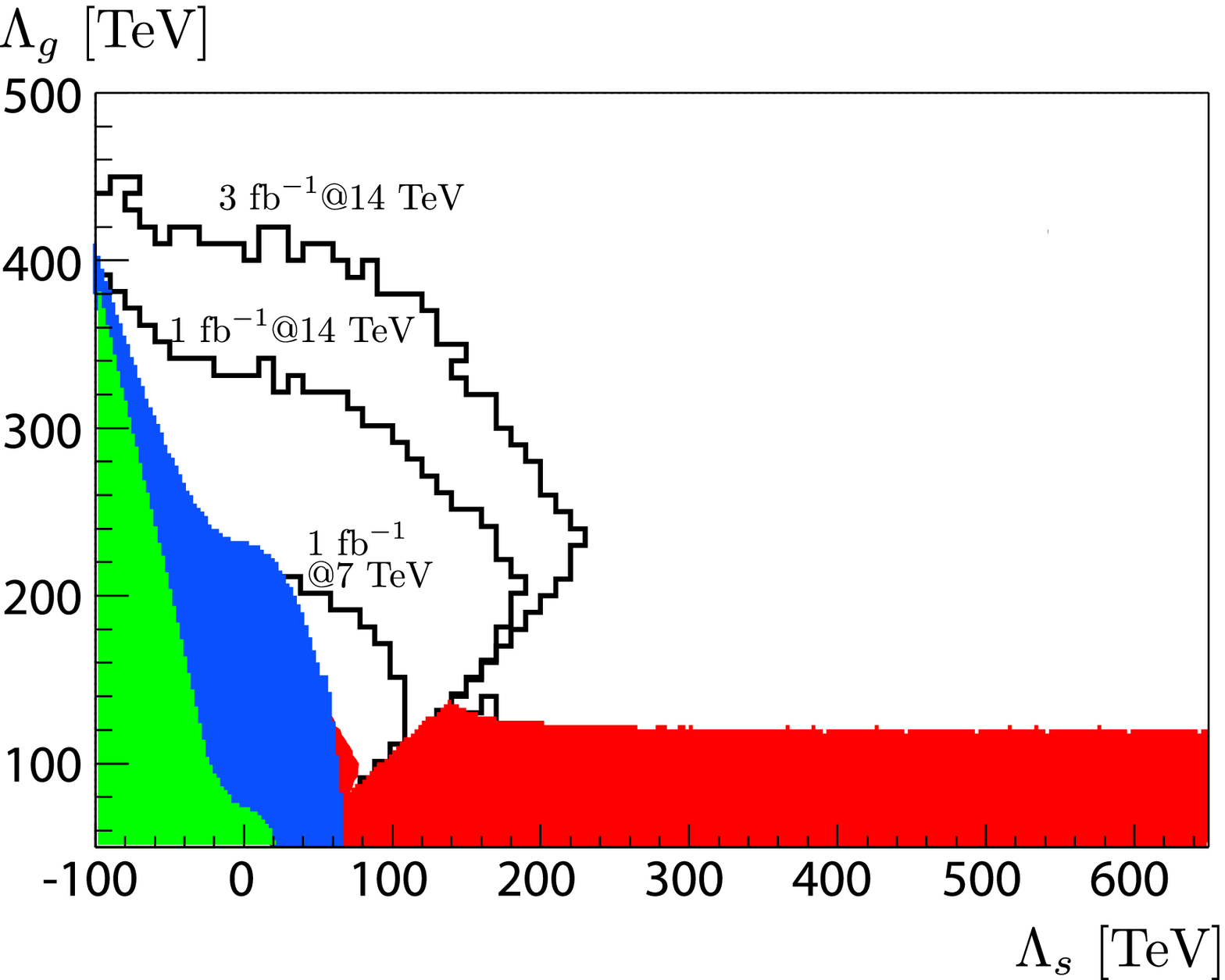}
}
\caption{LHC discovery region for $\tan\beta=10$. The red region is the region already excluded by the Tevatron 
trilepton \cite{Abazov:2009zi,Ruderman:2010kj} 
and photon signal search  \cite{Abazov:2010us}.
The blue region is the region  excluded by the LEP experiment \cite{Abdallah:2003xe}.
}
\label{fig:10d}
\end{figure}

\begin{figure}
\subfigure[2 photons]{
\includegraphics[clip, width=0.5\columnwidth]{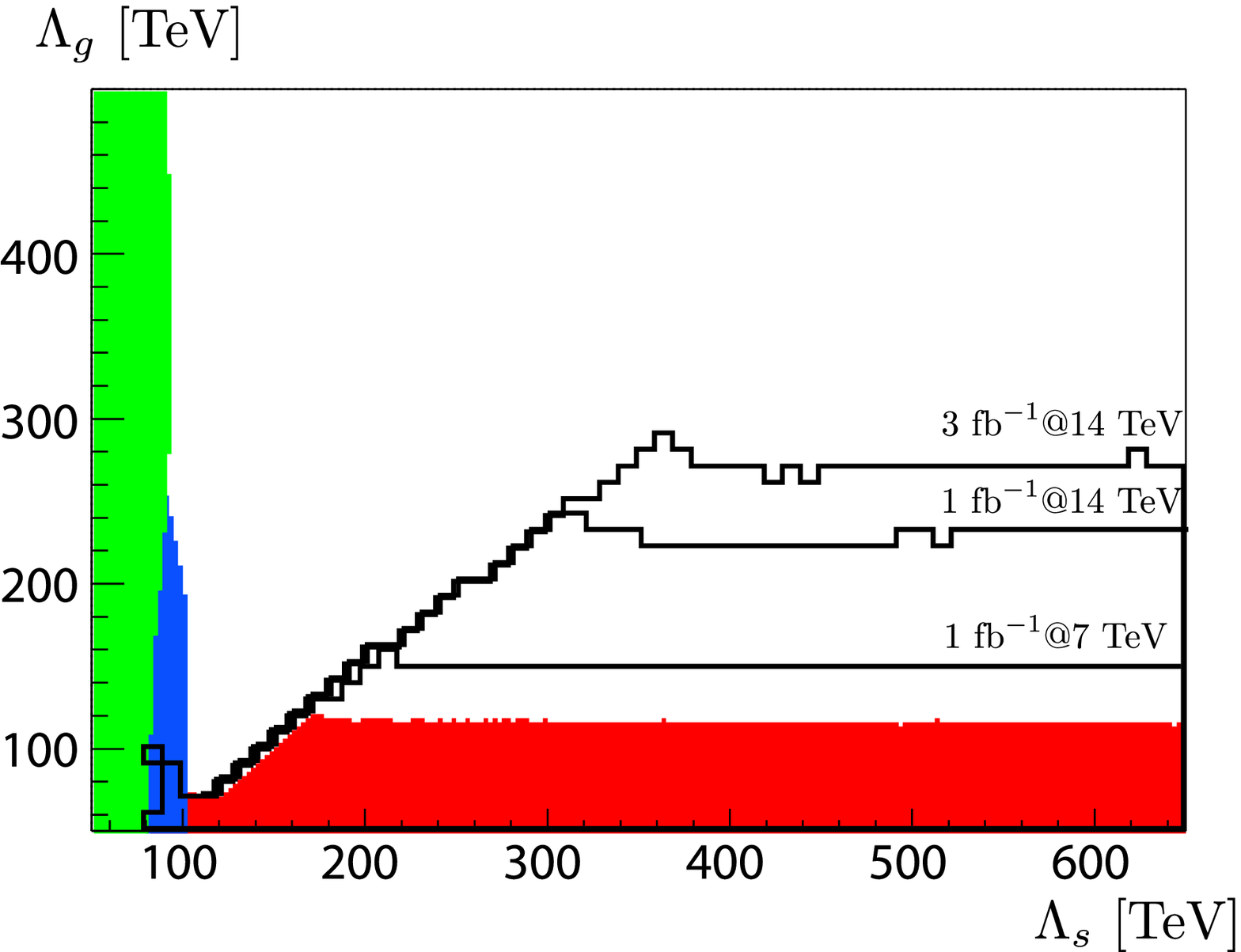}
}
\subfigure[3 leptons]{
\includegraphics[clip, width=0.5\columnwidth]{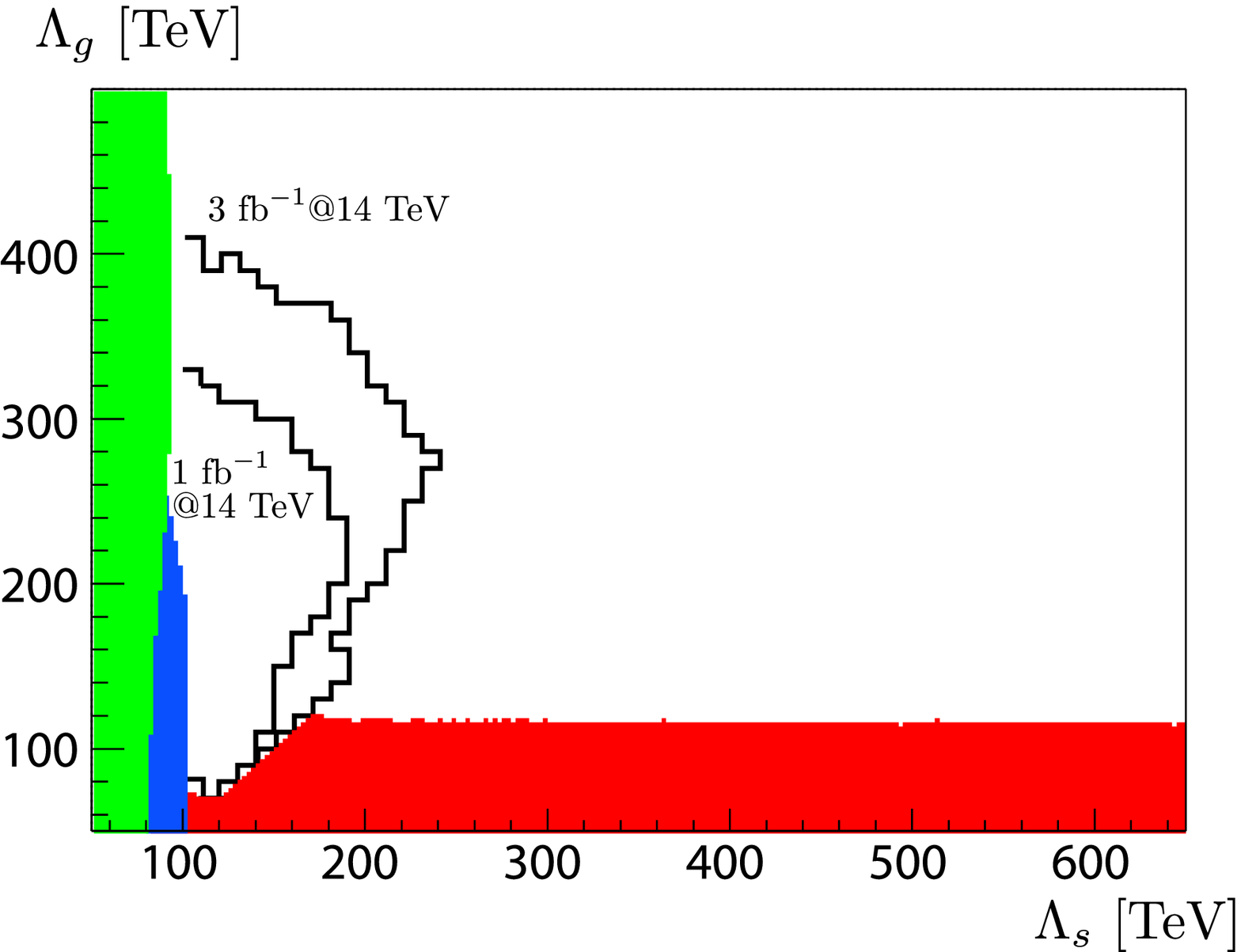}
}\\
\subfigure[4 or more leptons]{
\includegraphics[clip, width=0.5\columnwidth]{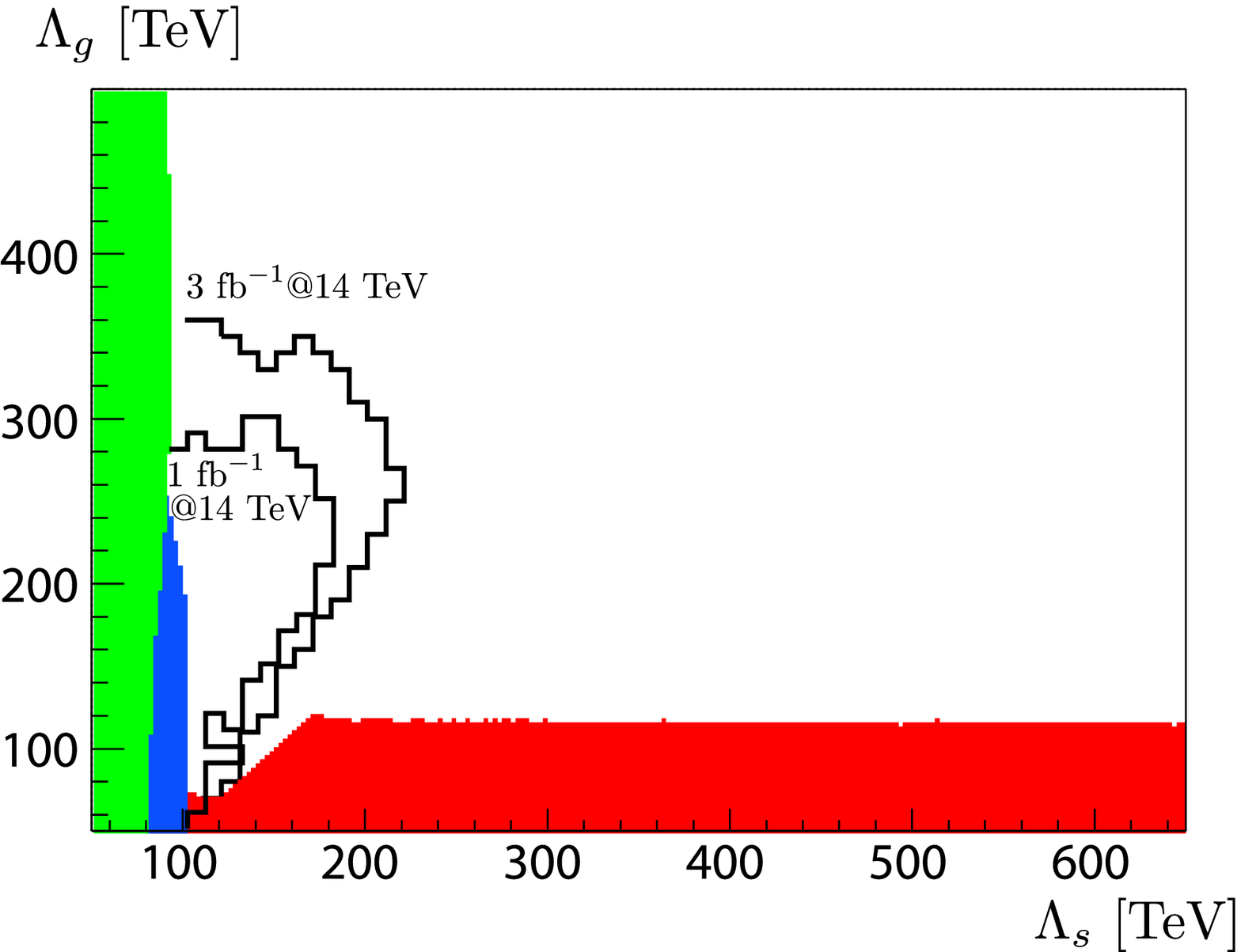}
}
\subfigure[Same-sign 2 leptons]{
\includegraphics[clip, width=0.5\columnwidth]{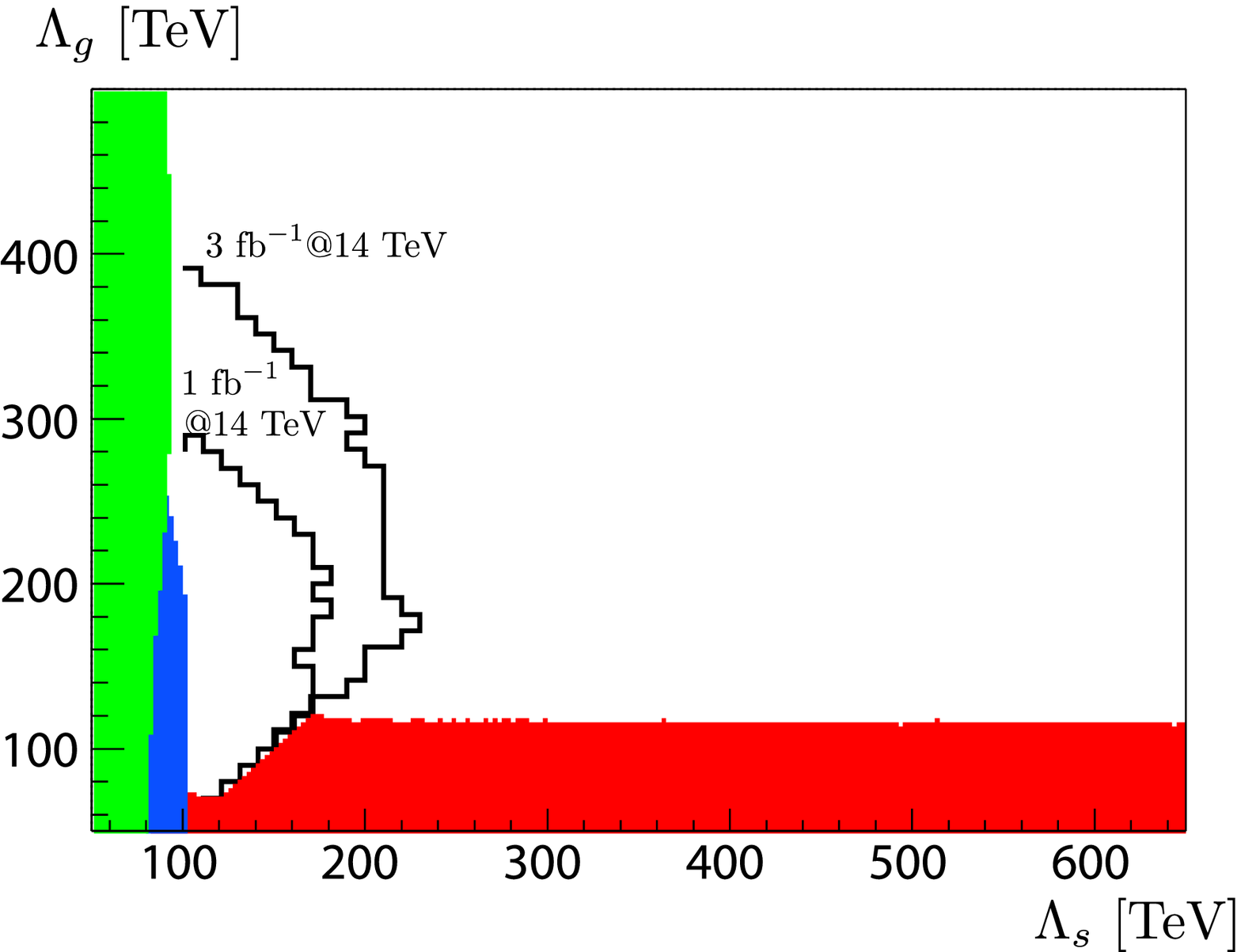}
}
\caption{Same as Fig. \ref{fig:10d}, but for $\tan\beta=40$.}
\label{fig:40d}
\end{figure}

In Figs. \ref{fig:10d} and \ref{fig:40d}, we show the discovery region of 
the gauge mediation model.


Among the searching modes we analyzed, the most important modes are
the two photon mode, same-sign two lepton mode, three lepton mode, and the four or more lepton mode.
The results for these modes are shown in the figures.
The red region is the region already excluded by the Tevatron trilepton \cite{Abazov:2009zi,Ruderman:2010kj} 
and photon signal search  \cite{Abazov:2010us}.
The blue region is the region  excluded by the LEP experiment \cite{Abdallah:2003xe}.

In the figures, we show the discovery region for 1 ${\rm fb}^{-1}$ at the 7 TeV run
and 1 and 3 ${\rm fb}^{-1}$ at the 14 TeV run.
First of all, we see a clear separation between the discovery regions of 2 photon mode and multi-lepton modes.
As discussed above, this corresponds to a change of the NLSP shown in Fig. \ref{fig:mass}.
The result for the 2 photon + missing search is illustrated in the top left figures labeled with (a).
The discovery region is determined by the gaugino masses, especially the wino mass.

In the stau-NLSP region, the multi-lepton modes are effective.
The three displayed modes, same-sign two lepton, three lepton, and four or more lepton modes
have similar discovery regions, which are roughly determined by the mass of squarks,
whose production is the dominant SUSY production.
We should mention that other lepton modes, such as zero-lepton multi-tau mode
and one lepton + multi-tau mode have narrower discovery regions than the above multi-lepton modes,
even in the ${\rm tan}\,\beta=40$ case. 
This is because a small fraction of produced tau leptons are detected as tau jets
because of its low identification efficiency and there are comparable amount of
$e/\mu$ leptons which come from the decay of the tau leptons.
Thus even for high tan$\,\beta$ cases, the multi-lepton modes are more important
than the less-lepton $+$ multi-tau modes.

\begin{figure}
\subfigure[${\rm tan}\,\beta=10$]{
\includegraphics[clip, width=0.5\columnwidth]{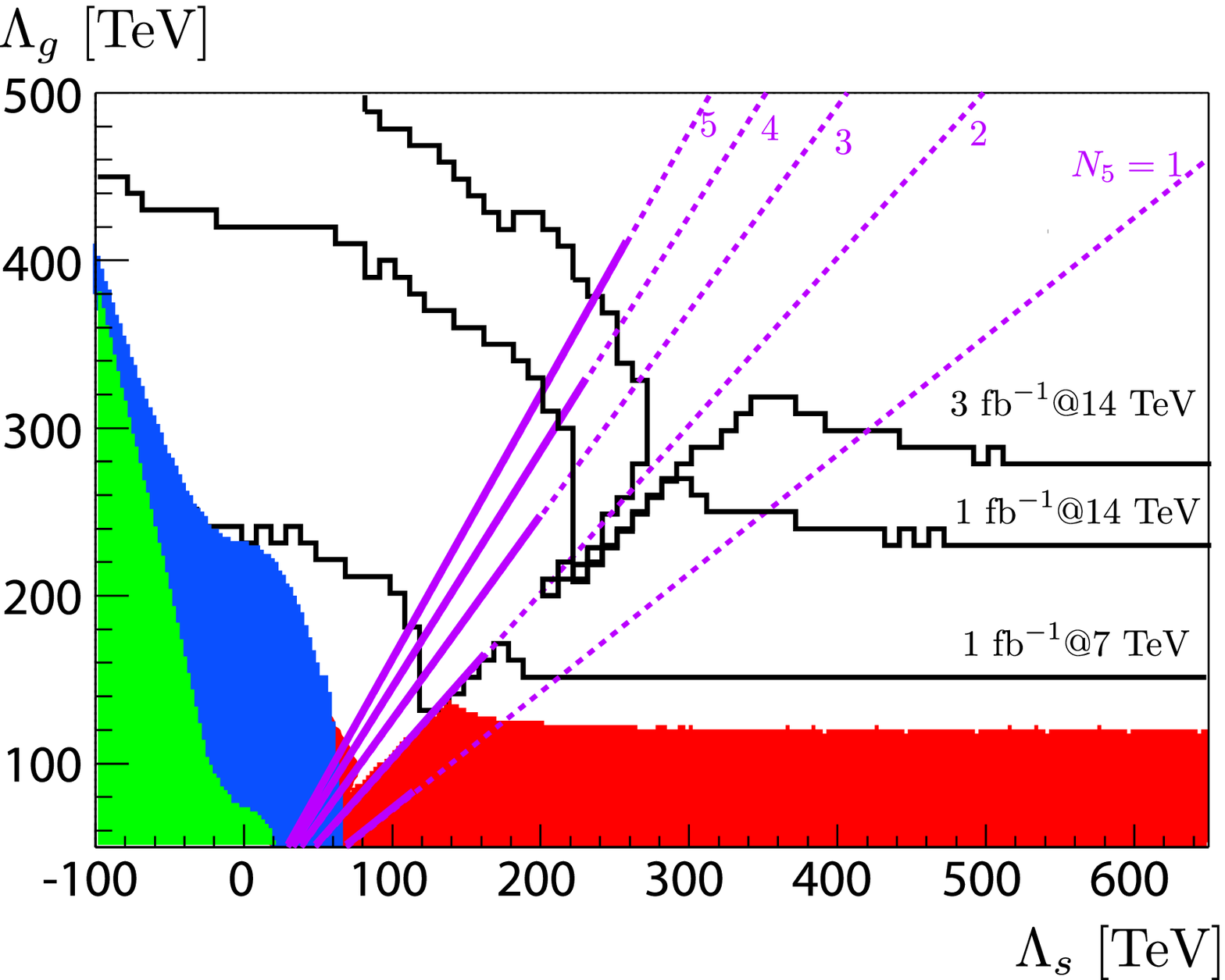}
}
\subfigure[${\rm tan}\,\beta=40$]{
\includegraphics[clip, width=0.5\columnwidth]{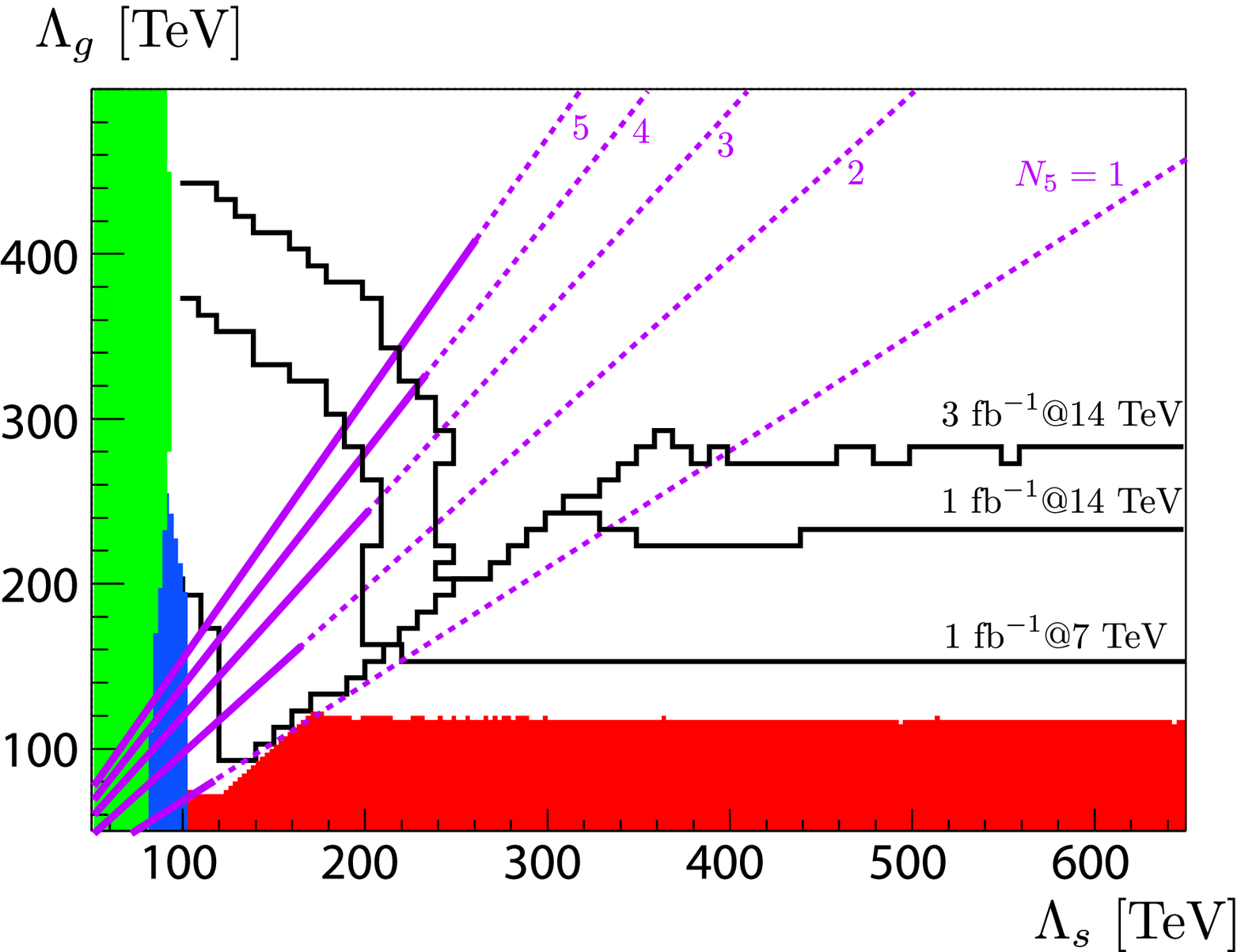}
}
\caption{Combined discovery region $\tan\beta=10$ and 40.
The red and blue regions are same as in Figs. \ref{fig:10d} and \ref{fig:40d}.
The purple lines show the points realized in the minimal GMSB with the number of messengers
$N_5=1$ to 5. On each line, the bold line shows the region with $\Lambda<80$ TeV (see Sec. \ref{minimal}).}
\label{fig:combine}
\end{figure}

Finally in Fig. \ref{fig:combine}, we show the combined result for the discovery region.
Discovery regions for all modes listed in Sec. \ref{modes} are combined.
In the figures, we also illustrate with purple lines, which correspond to the
minimal gauge mediation models with the number of messengers $N_5=1$ to 5.

\subsection{Comments on More General GMSB Models}\label{general}
\begin{figure}
\begin{center}
\includegraphics[clip, scale=0.6]{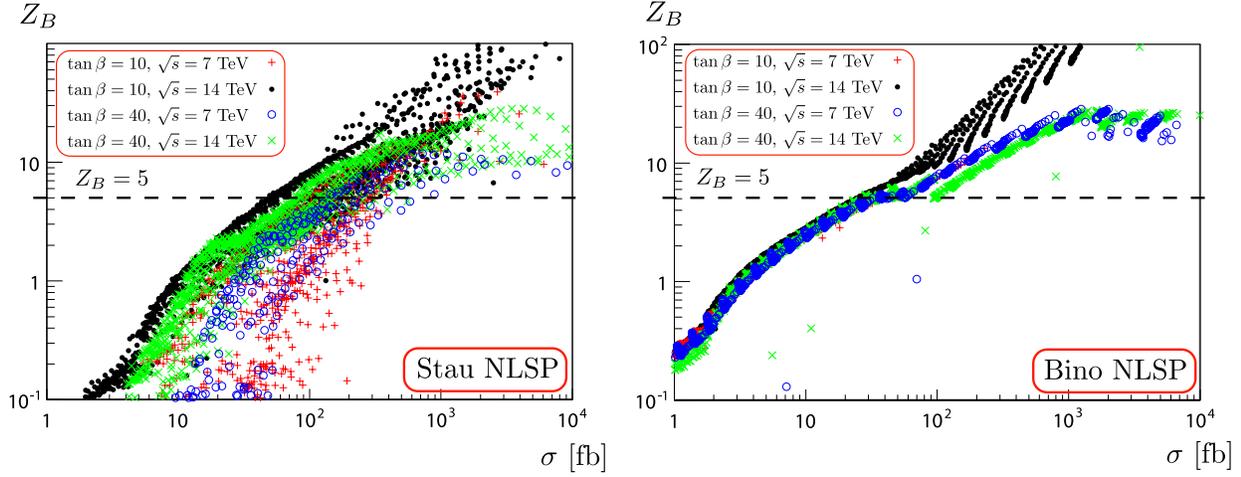}
\caption{Relation between the significance $Z_B$ and cross section $\sigma$ for the integrated luminosity $1 ~{\rm fb}^{-1}$. Left: the stau NLSP case, right: the bino NLSP case.
}
\label{fig:Z}
\end{center}
\end{figure}

We have assumed a ``GUT relation" for the gaugino and scalar masses.
However, in general GMSB models, this relation is not always maintained.
Although our present analysis cannot be directly applied to such models, 
the story does not change significantly, as long as the slepton or bino-like neutralino is the NLSP.
This is because although details of the SUSY signal depend on each SUSY spectrum,
the SUSY signal with multi-leptons + missing energy or multi-photon + missing energy is naturally expected in the
case of the slepton or neutralino NLSP for any SUSY mass spectrum.
Roughly speaking, if the number of SUSY events is ${\cal O}(10-100)$, SUSY can be discovered, as
shown in Fig.~\ref{fig:Z}.
In this figure, we show the scatter plot of the total SUSY production cross section $\sigma$ 
and the significance $Z_B$ for the integrated luminosity $1 ~{\rm fb}^{-1}$, assuming the ``GUT relation" for the gaugino and scalar masses.
One can see that in the bino NLSP case, almost all the region where the event number exceeds about 20-40 can be discovered.
This is because, in the bino NLSP case, the discovery relies almost only on diphoton + missing energy and
other objects such as jets or leptons are irrelevant.
In contrast, in the stau NLSP case, the required number varies widely, ${\cal O}(10-100)$.
In some parameter points, the SUSY cascade decays tend to emit tau-lepton instead of $e/\mu$ leptons, or
emissions of high-energy leptons are suppressed because of kinematical reasons.
In such cases, the required number of SUSY productions becomes large.
In contrast, if the SUSY spectrum prefers high-energy $e/\mu$ lepton emissions, 
the required number can be less than 50.
It is expected that this argument can be applied to more general GMSB models except for
specially-tuned parameter points, as long as the slepton or neutralino is the NLSP.
Thus we expect the SUSY particles can be discovered if the number of the SUSY particles reach ${\cal O}(10-100)$.

\begin{figure}
\begin{center}
\includegraphics[clip, scale=0.5]{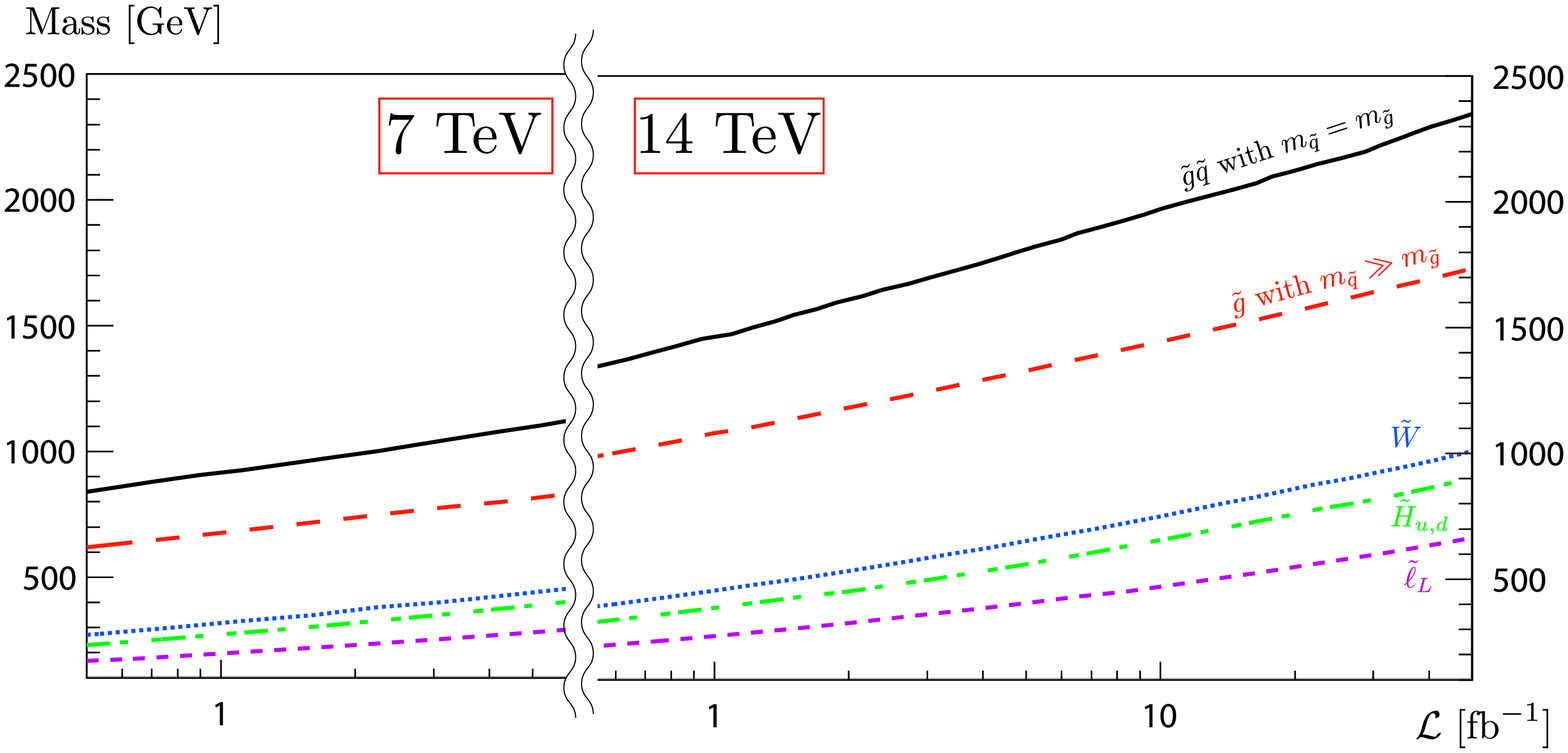}
\caption{Reachable SUSY masses at the LHC as a function of the integrated luminosity. Assuming that 100 SUSY events are required
for discovery, the lines show reachable SUSY masses for a given integrated luminosity. The left-hand side is for the 7 TeV run
and the right-hand side is for the 14 TeV run.}
\label{fig:general}
\end{center}
\end{figure}

In Fig. \ref{fig:general}, we show the masses of SUSY particles for which 100 pairs of SUSY particles can be produced for a given integrated luminosity.
If the required number of SUSY events for discovery is 100, as discussed above, the lines in  Fig. \ref{fig:general}
corresponds to the reachable mass range.

\section{Implications for Minimal GMSB Models}\label{minimal}

In the previous sections, we have analyzed the LHC signatures and discovery potential
of the LHC in a setup more general than the minimal GMSB models.
In this section, we discuss implications of our analysis for the minimal GMSB models.

Although, as described in the introduction, low-scale GMSB models are very attractive,
they tend to suffer from a serious problem: instability of the SUSY-breaking vacuum.
This problem stems from the fact that once one introduces the messenger particle in the theory for mediating the
SUSY-breaking effect, a supersymmetric vacuum, in which the messengers develop the VEV,
often comes into the theory and the former SUSY-breaking vacuum then becomes only a metastable vacuum.
Two possibilities can be considered to avoid this unwanted problem. One is to force
the metastable vacuum to have a lifetime much longer than the age of the universe, and the other is
to construct a model with a stable SUSY-breaking vacuum.

However, the mass scale of the SUSY particles is strongly constrained in both cases.
It is known that if the SUSY-breaking vacuum is stable, 
the gaugino masses are suppressed \cite{Komargodski:2009jf,Shirai:2010rr}.
In GMSB models with a perturbatively calculable stable vacuum, there is very strong 
upper bounds on the gaugino masses \cite{Sato:2009dk}.
Combining the updated null-result of diphoton+missing energy search at the Tevatron \cite{Abazov:2010us},
such types of GMSB models are completely excluded if $m_{3/2}<16$ eV.
The remaining possibilities are GMSB models with a metastable vacuum or 
strongly interacting GMSB models with a stable vacuum \cite{Hamaguchi:2008yu,Ibe:2010jb}.
In the following discussion, we concentrate on the case with a metastable vacuum.

There is also a strong constraint for the MSSM mass spectrum in this case \cite{Hisano:2007gb}.
To achieve a SUSY-breaking vacuum, whose lifetime is longer than the age of the universe, 
there is an upper bound on $\Lambda_g$ and $\Lambda_s$.
For example, if the number of messengers $N_5=1$ and if
we adopt the IYIT model \cite{IYIT} with an SP(1) gauge theory as the SUSY-breaking sector,
then $\Lambda_g$ must satisfy $\Lambda_g\lsim 80$ TeV \cite{Hisano:2007gb}.

This upper bound on $\Lambda_g$ (and therefore $\Lambda_s$) imposes severe constrains on
the MSSM mass spectrum. For example, if the number of messengers is one and $\tan\beta$ is not so large, then the NLSP is the neutralino
and the region $\Lambda_g<124$ TeV has been already excluded by the Tevatron diphoton+missing search \cite{Abazov:2010us}
as shown in Fig. \ref{fig:combine} (a).

One possible way to evade the above experimental bound is to increase the number of messengers $N_5$.
However, there is also an upper bound on the number of messengers $N_5$ to realize a successful GUT unification,
which is an important motivation for introducing low-energy supersymmetry.
A straightforward one-loop calculation yields the condition $N_5\leq5$.
According to a more detailed analysis in Ref.~\cite{Jones:2008ib}, the necessary condition
further strengthens to $N_5\leq4$ for $M={\cal O}(100)$ TeV,
otherwise the predictable GUT unification is spoiled.
Combining this constraint with the one resulting from the metastability condition,
we have constraints on the scales $\Lambda_g \lsim 320$ TeV and $\Lambda_s\lsim 230$ TeV.

In Fig. \ref{fig:combine}, the purple lines indicate the minimal GMSB models with $N_5=1$ to 5.
On each line for $N_5$, the bold region corresponds to the region with $\Lambda_g/N_5<80$ TeV,
where the condition from the stability of the SUSY-breaking vacuum is satisfied.
We see that for $N_5\leq4$, the allowed region can be tested at an integrated luminosity
3 ${\rm fb}^{-1}$ at 14 TeV for both cases with ${\rm tan}\,\beta=10$ and 40.

\begin{figure}
\begin{center}
\includegraphics[clip, scale=0.5]{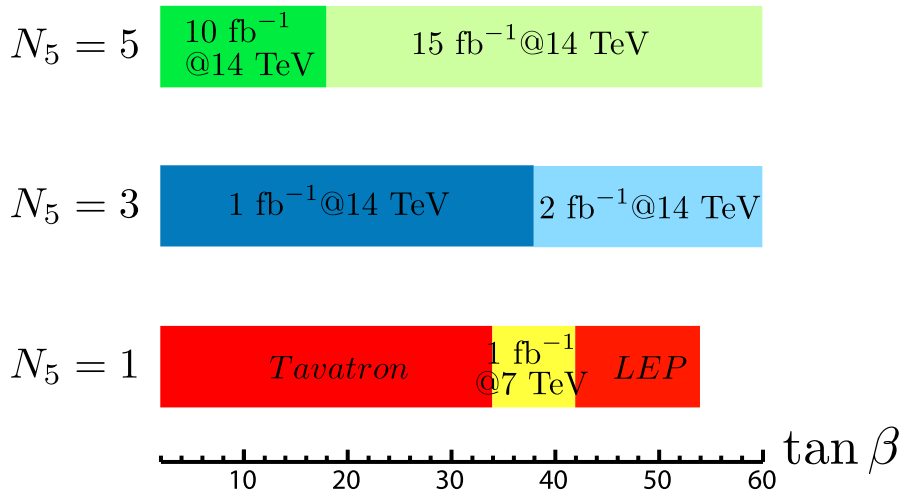}
\caption{Required luminosity for testing the minimal GMSB with $N_5=1,3,$ and 5 for varying $\tan\,\beta$.}
\label{fig:tanbeta}
\end{center}
\end{figure}

In Fig.~\ref{fig:tanbeta}, we show the integrated luminosity required to test the minimal GMSB models
with $\Lambda_g/N_5=80$ TeV for varying ${\rm tan}\,\beta$ and $N_5=1,3,$ and 5.
The NLSP is the stau except for the region ${\rm tan}\,\beta\lsim35$ with $N_5=1$, which is already
excluded by the Tevatron experiment. Therefore, the relevant discovery mode is multi-lepton + missing energy.
We note that when $N_5=1$, almost all the regions in ${\rm tan}\,\beta$ are already excluded by
the current experimental limits, but a small region above ${\rm tan}\,\beta\gsim35$ survives.
This small region can be tested at a very early stage with 1 ${\rm fb}^{-1}$ at 7 TeV.
We also see that even for $N_5=5$, all the regions in ${\rm tan}\,\beta$ can be tested
with an integrated luminosity of about 10 ${\rm fb}^{-1}$ at 14 TeV.
For larger $\tan\beta$, larger integrated luminosity is required to test the models.
The reason is that more tau leptons instead of $e/\mu$ leptons tend to be emitted in SUSY cascade decays
for larger ${\rm tan}\,\beta$, because of both kinematics and the ${\rm SU}(2)_L$ interaction carried
by the mixed stau.

\section{Conclusion and Discussion}
In this paper we have investigated the LHC discovery region for low-scale GMSB models
with a setup applicable for a wider class of models other than the minimal GMSB models.
We have performed a comprehensive study while giving careful treatment for realistic experimental setups.
Our result is thus a very conservative one.
Although we have set the lepton and photon detection efficiencies are 100 \% for the SM background,
if we impose the same fake rates as used for the SUSY signals, the reduction of the background is expected, which leads
to a wider discovery region.
In addition, we have used the leading-order SUSY production cross section.
Generally, the next-leading order cross section is larger than the leading-order one.
Thus, if we adopt the next-leading cross section, the discovery region is extended by about 10 \%.

We have applied our result to the minimal GMSB models, in which the MSSM SUSY-particle masses
have strong upper bounds from the stability condition of the SUSY-breaking vacuum.
We have shown that all the region can be tested at an early stage of the LHC if $m_{3/2}\lsim16$ eV.
Let us comment on the case with a lighter gravitino.
Approximately, the upper bound of the scales $\Lambda_g$ and $\Lambda_s$ is proportional to $\sqrt{m_{3/2}}$.
If in the future the upper bound on the gravitino mass will be improved to, say, 3 eV,
then it will be predicted that the SUSY can be discovered at a very early stage $\sqrt{s}=7$ TeV, even if $N_5=5$.

\section*{Acknowledgements}
We are grateful to K. Hamaguchi and T. Yanagida for careful reading of the manuscript and useful
comments.
This work is supported in part by JSPS
Research Fellowships for Young Scientists and by
World Premier International Research Center Initiative, MEXT, Japan.

\end{document}